\documentclass[10pt,a4paper,reqno]{amsart}
\usepackage{parskip, color}
\usepackage[mathscr]{euscript}
\newtheorem{theorem}{\bf Theorem}[section]
\newtheorem{remark}[theorem]{Remark}
\newtheorem{prop}[theorem]{Proposition}
\renewcommand{\eqref}[1]{(\ref{#1})}
\usepackage{xcolor}
\usepackage[colorlinks=true,linkcolor=blue,citecolor=orange]{hyperref}

\begin{document}

\title[Phase recovery for discrete Schr\"{o}dinger operators]{Phase recovery from phaseless scattering data for discrete Schr\"{o}dinger operators}

\author{Roman Novikov}
\address{Roman G. Novikov, CMAP, CNRS, \'{E}cole polytechnique, Institut Polytechnique de Paris, 91128 Palaiseau, France \newline
\& IEPT RAS, 117997 Moscow, Russia
}
\email{novikov@cmap.polytechnique.fr}

\author{Basant Lal Sharma}
\address{Basant Lal Sharma, Department of Mechanical Engineering, Indian Institute of Technology Kanpur, Kanpur, 208016 UP, India
}
\email[Corresponding author]{bls@iitk.ac.in}

\begin{abstract}
We consider scattering for the discrete Schr\"{o}dinger operator on the square lattice $\mathbb{Z}^d$, $d \ge 1$, with compactly supported potential. We give formulas for finding the phased scattering amplitude from phaseless near-field scattering data.

\end{abstract}

\maketitle 

\section{Introduction}

We consider the discrete Schr\"{o}dinger equation
\begin{equation}
\Delta \psi(x)+v(x)\psi(x)=E\psi(x),\quad x\in\mathbb{Z}^{d}, d\ge 1.
\label{discschro}
\end{equation}
We assume that
$\Delta$ is the discrete Laplacian defined by
\begin{equation}
\Delta\psi\left(x\right) =\sum_{|x^{\prime}-x|=1}
\psi\left(x^{\prime}\right), \quad x, x'\in\mathbb{Z}^{d},
\label{discLap}
\end{equation}
$v$ is a scalar potential
such that
\begin{equation}
\textrm{supp }v\subset \mathcal{D},
\label{defv}
\end{equation}
where $\mathcal{D}$ is bounded in $\mathbb{Z}^{d}$, 
and 
\begin{equation}
E\in S:=
\left[ -2d,2d\right] \backslash S_{0},
\label{defS}
\end{equation}
where
\begin{eqnarray}
S_{0}&:=&\{ \pm4n\textrm{ when }d\textrm{ is even,}\nonumber\\
&&\qquad\pm2(2n+1)\textrm{ when }d\textrm{ is odd, }n\in\mathbb{Z}\textrm{ and }2n\leq d\}.
\label{defS0}
\end{eqnarray}

The operator $H = \Delta + v$, where $\Delta$ is defined by \eqref{discLap} for complex valued $\psi$ and $v$ is the multiplication operator for functions $\psi$ on $\mathbb{Z}^{d}$, is our discrete Schrödinger operator.

The discrete Schr\"{o}dinger equation \eqref{discschro} appears in the TB (tight-binding) model of the electrons in crystals, as in many cases the electrons of a crystal are strongly attached to the atoms 
\cite{Bloch,Slater} involving a very weak hopping interaction with the neighboring atoms 
due to quantum tunnelling. The TB model is also closely related to LCAO (linear combination of atomic orbitals) 
\cite{Harrison}.
Moreover, for example when $d=1$ and $d=2$, it has played a crucial role in uncovering significant
phenomena associated with electrons in crystals
\cite{Anderson,Economou,Andoqm}.
At a different length scale, in the domain of electrical engineering, lattice structures of LC circuits involve similar difference equations
where they play an important role in network synthesis and filter design \cite{Kuo}; lately such equations also appear in the lumped circuit models for electromagnetic metamaterials \cite{Itoh}. 
The same equation also appears, for $d=1,2$, in case of of time harmonic lattice waves \cite{Brillouin,Maradudin} and reveals structure in simple cases also for $d=3$. In particular, the case $d=2$ corresponds to a discrete analogue of anti-plane shear waves in elastic continuum. Examples of forward analysis of such equations, in the case $d=2$, with an exact solution of scattering of time harmonic lattice waves by atomically sharp crack tips and rigid constraints, can be found in \cite{Bls0,Bls2,Bls1,Bls3,BlsMaurya1,BlsMaurya2,BlsMaurya3,BlsMis}.
The physical literature concerning the discrete Schr\"{o}dinger equation also includes, in particular, \cite{Zakhariev}.

The discrete Schr\"{o}dinger equation was studied by \cite{Eskina1}, \cite{Eskina2}, \cite{Shaban} and \cite{Isozaki1} from pure mathematical viewpoint.

Let
\begin{equation}
\Gamma(E)=\{k:k\in T^{d},\textrm{ }\phi\left(k\right) =E\},\quad E\in\left[ -2d,2d\right].
\label{defGamma}
\end{equation}
Here $T^{d}=\mathbb{R}^{d}/2\pi\mathbb{Z}^{d}$
and
\begin{equation}
\phi\left(k\right) =2{{\sum_{i=1}^{d}}}\cos k_{i}.
\label{defphi}
\end{equation}

For Eq. \eqref{discschro} we consider the scattering solutions 
\begin{equation}
\psi^+(x, k)=\psi_{0}^+(x, k)+\psi_{sc}^+(x, k),
\label{psiplus}
\end{equation}
where 
\begin{equation}
\psi_{0}^+(x, k)=e^{i {{k}} \cdot x},\quad {{k}}\in \Gamma(E),\quad x\in\mathbb{Z}^{d},
\label{discinc}
\end{equation}
and $\psi_{sc}^+(x, k)$ is the outgoing solution for the non-homogenous equation
\begin{equation}
\Delta \psi_{sc}-E\psi_{sc}+v\psi_{sc}=-v\psi_{0}^+,
\label{discschro2}
\end{equation}
obtained using the limiting absorption principle; see \cite{Shaban}.
Strictly speaking, to consider $\psi^+$ we also assume that $v$ is real-valued or that energy $E$ is not singular for the case of complex $v$.
Note also that $\psi^+_0$ solves $\Delta \psi^+_0=E \psi^+_0$ in $\mathbb{Z}^{d}$, in view of \eqref{defGamma}, \eqref{defphi}, and \eqref{discinc}.

If 
\begin{equation}
2d-4<\left| {{E}}\right| <2d,
\label{assumeE}
\end{equation}
then the surface
$\Gamma({{E}})$ is smooth strictly convex with non-zero principal curvatures, and there is a unique point
\begin{equation}
{\gamma}={\gamma}\left( \omega, {{E}}\right) \in\Gamma(E),\quad \omega\in S^{d-1},
\label{defkout}
\end{equation}
where the outward normal 
to the surface $\Gamma(E)$ is equal to
$\omega.$ 

\begin{remark}
For positive $E$, it is convenient to consider $\Gamma(E)$ to be symmetric with respect to the origin $\mathtt{O}$ in $\mathbb{R}^d$.
For negative $E$, it may be convenient to consider $\Gamma(E)$ to be symmetric with respect to the point $\mathtt{O}_\pi$ in $\mathbb{R}^d$, where all coordinates are equal to $\pi$.    
In addition, when $2d-4<E<2d$ ($-2d <E<-2d+4$), the surface $\Gamma(E)$ is located strictly inside the cube $[-\pi, \pi]^d$ ($[0, 2\pi]^d$, respectively); see Lemma 2 in \cite{Shaban}.
\label{remark1p1}
\end{remark}

Under assumption \eqref{assumeE}, 
$\psi^{+}$ has the asymptotic expansion
\begin{eqnarray}
\psi^{+}(x, k) &=e^{i{{k}}\cdot{{x}}}+\dfrac{e^{+i\mu\left(\omega, {{E}}\right) \left| {{x}}\right| }
}{\left| {{x}}\right| ^{\frac{d-1}2}}f^{+}\left({{k}},\omega\right)
+O(\dfrac1{\left| {{x}}\right| ^{\frac{d+1}2}})\nonumber \\
&\qquad\quad\textrm{as
}\left| {{x}}\right| \longrightarrow\infty,
\quad \omega=\frac{x}{|x|}, \quad x\in\mathbb{Z}^d,
\label{asympsiplus}
\end{eqnarray}
where
\begin{equation}
\mu\left(\omega, {{E}}\right)={\gamma}(\omega, {{E}})\cdot\omega,
\label{defmu}
\end{equation}
and the coefficient $f^{+}\left({k},\omega\right) $
is smooth and the remainder can be estimated uniformly in
$\omega$
\cite{Eskina1,Eskina2,Shaban}.
The coefficient $f^{+}$ is the scattering amplitude for \eqref{discschro}.
In many respects, expansion \eqref{asympsiplus} is similar to the related expansion for the case of continuous Schr\"{o}dinger equation, 
where 
\begin{equation}
\Gamma(E)=S^{d-1}_{\sqrt{E}}=\{k\in\mathbb{R}^d: |k|=\sqrt{E}\}, \quad \mu(\omega, {{E}})=\sqrt{E}>0;
\label{defcontE}
\end{equation}
see, for example, \cite{Novikovrev}.
For both discrete and continuous cases, the remainder $O(|x|^{-(d+1)/2})\equiv0$ for $d=1$.

\begin{remark}
    Note that in \eqref{asympsiplus}, \eqref{defmu}, we have 
\begin{eqnarray}
&&\mu(\omega, E)|x|=\gamma(\omega, E)\cdot x,\\ 
&&\mu(\omega, E)>0\textrm{ if }2d-4<E<2d, \label{defnmuposE}\\
&&\mu(\omega, E)=\mathtt{O}_\pi\cdot \omega+\mu_-(\omega, E)\textrm{ if }-2d<E<-2d+4, \nonumber\\
&&\qquad\qquad\qquad\qquad\qquad\textrm{ where } \mu_-(\omega, E)>0,\qquad
\end{eqnarray}
and $\mathtt{O}_\pi$ is defined in Remark \ref{remark1p1}.
In addition, $\mu$ in \eqref{defnmuposE} differs in sign from that defined in \cite{Shaban}.
\end{remark}

In quantum mechanics, according to the Born principle, the complex values of wave functions $\psi^+$ and scattering amplitude $f^+$ do not have direct physical interpretations, whereas the absolute values of these functions admit probabilistic interpretation and can be measured directly in physical experiments.
For example, in the problem of electronic transport through interfaces, naturally described in terms of transmission and reflection, following the Landauer-B\"{u}ttiker approach
\cite{Landauer1,Buttiker1}, the scattering amplitudes (the probability amplitude) decide the conductance in the linear response regime; see, for example, \cite{Blsbif,Blshexa,Blshexa2,Blsstep} for an application to transport in waveguides where the forward problem of discrete Schr\"{o}dinger equation has been solved exactly.

In the present work, we give explicit asymptotic formulas for finding complex $f^+$ from $|\psi^+|^2$ measured on $\mathbb{Z}^d\setminus \mathcal{D}$, under assumption \eqref{assumeE}.
In many respects these formulas are similar to the formulas in \cite{Novikov5,Novikov4,Novikov1d} given for the case of continuous Schr\"{o}dinger equation.
In particular, for equation \eqref{discschro}, these formulas can be used in the framework of phaseless inverse scattering from 
$|\psi^+|^2$, using results on inverse scattering from $f^+$ with phase information.
For equation \eqref{discschro}, some results on inverse scattering from $f^+$, in fact, are given in \cite{Zakhariev,Eskina2,Isozaki1,Isozaki2,Tarnopol}.
In connection with inverse scattering for continuous Schr\"{o}dinger equation, see, for example, the review article \cite{Novikovrev} and references therein.

The present article can be considered as the first work on phaseless inverse scattering for the
discrete Schr\"{o}dinger equation \eqref{discschro}.
In connection with phaseless inverse scattering for the
continuous Schr\"{o}dinger equation and other continuous equations of wave propagation, see, for example, \cite{Chadan,Aktosun,Hohage,Klibanov1,Klibanov2,Klibanov3,Romanov,Novikov4,Novikovrev,Novikov2,Sivkin} and references therein.

The main results of the present article are given in Section \ref{mainresult} and proved in Sections \ref{proofsec3} and \ref{proofsec4}.

\section{Main Results}
\label{mainresult}
Let us define
\begin{equation}
a({{x}}, {{k}})=\left| {{x}}\right| ^{\frac{d-1}2}(|\psi^{+}\left({{x}}, {k}\right)|^2 -1), \quad {x}\in \mathbb{Z}^d\setminus\{0\}, \quad {k}\in\Gamma(E),
\label{defaxk}
\end{equation}
where $\psi^{+}\left({{x}}, {k}\right)$ is defined as in \eqref{psiplus} in introduction.

In particular, we consider $a({{x}}, {{k}})$ at two measurement points 
\begin{eqnarray}
x=\textrm{Int}(s\omega)\textrm{ and }y=x+\zeta,\quad
s>0, \quad
 \omega \in S^{d-1},\quad \zeta\in \mathbb{Z}^d\setminus\{0\},
\label{defxy}
\end{eqnarray}
where 
\begin{equation}
\textrm{Int}(\xi)=\sum_{i=1}^d~\textrm{sgn}({\xi_i})~ \lfloor |\xi_i|\rfloor~\mathbf{e}_i, \quad \xi\in\mathbb{R}^d,
\label{defintxi}
\end{equation}
and $\lfloor\cdot\rfloor$ denotes the floor function while $\mathbf{e}_i$ are unit basis vectors in $\mathbb{Z}^d$ or in $\mathbb{R}^d$.

We use the
notation
\begin{equation}
\hat{x}=x/|x|,\quad
\hat{y}=y/|y|,,\quad x, y\in \mathbb{Z}^d.
\label{defnhat}
\end{equation}
Note that $\hat{x}\ne\omega$, in general, for $x$ in (20). Note also that $\hat{x}$ and $\hat{y}$ in \eqref{defnhat} are not arbitrary points of $S^{d-1}$ because of the discreteness of $x$ and $y$. More precisely, $\hat{x}$ and $\hat{y}$ belong to a countable, everywhere dense subset of $S^{d-1}$.

In dimension $d\ge2$, we have, in particular, the following theorem.
\begin{theorem}
Suppose that assumptions \eqref{defv} and \eqref{assumeE} are satisfied, $d\ge 2$, and definitions \eqref{defkout},
\eqref{defmu}
hold.
Then we have the following formulas:
\begin{eqnarray}
f^+(k, \omega)=\frac{1}{D}\big(e^{i({{k}}\cdot{{y}}-\mu\left(\hat{y}, {{E}}\right) \left| y\right| )} a(x, {k}) - e^{i({{k}}\cdot{{x}}-\mu\left(\hat{x}, {{E}}\right) \left| x\right| )}a(y, {k})+O(s^{-\sigma})\big)\nonumber\\
\qquad \textrm{ as }s\to+\infty,
\label{solfplus}\\
D=2i \sin\big({{k}}\cdot \zeta+\mu(\hat{x}, {{E}})|x|-\mu(\hat{y}, {{E}})|y|\big),\nonumber\\
\quad \sigma = 1/2 \textrm{ for }d = 2, \quad \sigma = 1 \textrm{ for } d \ge 3,\nonumber
\end{eqnarray}
where ${k}\in\Gamma(E)$, $\omega\in S^{d-1}$, $a(x, {k})$ is defined by \eqref{defaxk},
and $x, y, \zeta, s$ are as in \eqref{defxy}.
\label{thm2p1}
\end{theorem}

We consider \eqref{solfplus} assuming that $D\ne0$. In addition,
we use the following formula:
\begin{eqnarray}
D
&=2i\sin \big(({{k}}-\gamma(\omega, {{E}}))\cdot\zeta+O(s^{-1})\big)\textrm{ as }s\to+\infty,
\label{Dexp}
\end{eqnarray}
uniformly in $\omega$ for fixed $\zeta$, where $\hat{x}=x/|x|.$

For fixed ${k}$ and $\zeta$, in view of formula \eqref{Dexp}, 
formula
\eqref{solfplus} can be used for finding $f^+$ under the condition that $\omega\in S^{d-1}\setminus \mathcal{E}_{{k},\zeta}$ 
with
\begin{equation}
\mathcal{E}_{{k},\zeta}=\{\omega\in S^{d-1}: ({{k}}-\gamma(\omega, {{E}}))\cdot\zeta=0~(\textrm{mod }\pi)\}.
\label{defEkg}
\end{equation}
In addition, 
\begin{equation}
\textrm{Meas }\mathcal{E}_{{k},\zeta}=0\textrm{ in }S^{d-1},
\label{measE}
\end{equation}
at least under assumption \eqref{assumeE}.

Theorem \ref{thm2p1} and formulas \eqref{Dexp} and \eqref{measE} are proved in Section \ref{proofsec3}.

In order to improve the remainder $O(s^{-1/2})$ in formula \eqref{solfplus} for $d=2$, we also give the following result.
\begin{prop}
Suppose that the assumptions of Theorem \ref{thm2p1} for $d=2$ hold,
and 
\begin{eqnarray}
f^+_0(k, \omega):=\frac{1}{D}\bigg(e^{i({{k}}\cdot{{y}}-\mu\left(\hat{y}, {{E}}\right) \left| y\right| )} a(x, {k}) - e^{i({{k}}\cdot{{x}}-\mu\left(\hat{x}, {{E}}\right) \left| x\right| )}a(y, {k})\bigg).
\label{deff0plus}
\end{eqnarray}
Then
\begin{eqnarray}
f^+(k, \omega)
&=&f^+_0
-\frac{1}{D}(e^{i({{k}}\cdot{{y}}-\mu\left(\hat{y}, {{E}}\right) \left| y\right| )}-e^{i({{k}}\cdot{{x}}-\mu\left(\hat{x}, {{E}}\right) \left| x\right| )})\frac{1}{s^{1/2}}\overline{f^{+}_0}f^{+}_0\nonumber\\
&&+\frac{1}{D^3}O(s^{-1}), \textrm{ as }s\to+\infty,\label{solfplusd2}
\end{eqnarray}
where $f^+_0=f^+_0(k, \omega)$.
\label{prop2p4n}
\end{prop}
Proposition \ref{prop2p4n} is proved in Section \ref{proofsec4n}.

\begin{remark}
    Formulas \eqref{solfplus}, \eqref{deff0plus}, \eqref{solfplusd2} are invariant
    with respect to the changes $\Gamma(E)\rightarrow \Gamma(E)+2\pi z,$ where $z\in\mathbb{Z}^d.$
    One can see this using the formulas
    $\mu\left(\hat{x}, {{E}}\right)|x|=\gamma\left(\hat{x}, {{E}}\right)\cdot x$, $\mu\left(\hat{y}, {{E}}\right)|y|=\gamma\left(\hat{y}, {{E}}\right)\cdot y$.
\end{remark}

\begin{remark}
In many respects, Theorem \ref{thm2p1} is similar to Theorem 3.1 in \cite{Novikov4} given for the continuous case.
However, in Theorem \ref{thm2p1}, 
the direction $\hat{y}$ is typically different from $\hat{x}$ in view of definitions \eqref{defxy}, although, asymptotically these directions become equal to $\omega$, as $s\to +\infty$.
We recall that $\hat{y}=\hat{x}$ in Theorem 3.1 in \cite{Novikov4}.
In turn, Proposition \ref{prop2p4n} is similar to Proposition 9.1 in \cite{Novikov2021} given for the continuous case.
\label{thm2p2}
\end{remark}

\begin{remark}
Formulas \eqref{solfplus}--\eqref{solfplusd2} can be also considered for the continuous case, assuming
\eqref{defcontE}
and assuming that $\zeta\in\mathbb{R}^d\setminus\{0\}$ in \eqref{defxy}.
These formulas are new even for the continuous case if $\hat{\zeta}\ne \hat{x}$.
\label{thm2p3}
\end{remark}

Let $\mathcal{D}_{\epsilon} = \{x \in\mathbb{Z}^d: \text{dist }(x, \mathcal{D}) \le \epsilon\}.$ Note that $\mathcal{D}_0 = \mathcal{D}$. Note also that \eqref{discschro}--\eqref{defv} reduces to \eqref{discschro}--\eqref{discLap} with $v \equiv 0$ on $\mathbb{Z}^d\setminus \mathcal{D}_{1}$, in view of the definition \eqref{discLap}.

Note that, for $d=1$,
\begin{equation}
\Gamma(E)=\{-\arccos\frac{E}{2}, \arccos\frac{E}{2}\},
\quad \mu(\omega, E)=\arccos\frac{E}{2},
\label{eqnk1d}
\end{equation}
where $\arccos(\kappa)\in[0,\pi]$ for $\kappa\in[-1, 1]$.
In this case, we consider $\Gamma(E)$ to be symmetric with respect to the origin in $\mathbb{R}$ even for negative $E$ inspite of Remark \ref{remark1p1}.

Let $\mathbb{R}^-$ denote the set of negative real numbers.

In dimension $d=1$, we have, in particular, the following propositions.
\begin{prop}
Suppose that assumption \eqref{defv} holds, $|E|<2, d=1$, and ${k}=\arccos({E}/{2})$.
Let $x, y\in (\mathbb{Z}\setminus\mathcal{D})\cap\mathbb{R}^-, x\ne y\textrm{ mod }(\pi/(2{k})).$
Then
\begin{eqnarray}
s_{21}&:=&f^+({k},-1)\nonumber\\
&=&\frac{1}{D}\big( e^{2iky}a(x, {k})-e^{2ikx}a(y, {k})+|s_{21}|^2(e^{2ikx}-e^{2iky})\big),\label{eqprop2p4}\\
D&=&2i\sin(2{k}(y-x)).\nonumber
\end{eqnarray}
\label{prop2p4}
\end{prop}

\begin{prop}
Suppose that assumption \eqref{defv} holds, $|E|<2, d=1$, and ${k}=\arccos({E}/{2})$.
Let $x_1, x_2, x_3\in (\mathbb{Z}\setminus\mathcal{D})\cap\mathbb{R}^-,$
and $x_i\ne x_j\textrm{ mod }(\pi/{k})$.
Then
\begin{eqnarray}
s_{21}&=&\frac{1}{D}\big( 
(e^{2ikx_3}-e^{2ikx_1})(a(x_2, {k})-a(x_1, {k}))\nonumber\\
&&\qquad+(e^{2ikx_1}-e^{2ikx_2})(a(x_3, {k})-a(x_1, {k}))\big),\label{eqprop2p5}\\
D&=&16i(\sin(k(x_2 - x_3)) \sin(k(x_2 - x_1)) \sin(k(x_1 - x_3))).\nonumber
\end{eqnarray}
\label{prop2p5}
\end{prop}

In these propositions, $a(x, k)$ is defined by \eqref{defaxk} for $d=1$.

Propositions \ref{prop2p4} and \ref{prop2p5} are proved in Section \ref{proofsec4}.

Formulas similar to \eqref{eqprop2p4}, \eqref{eqprop2p5} can be also given for $s_{12}:=f^+(k,1)$, where ${k}=-\arccos({E}/{2}).$

\begin{remark}
In fact, propositions \ref{prop2p4} and \ref{prop2p5} are completely similar to theorems 2.1 and 2.2 in the arXiv preprint of \cite{Novikov1d} given for the continuous case.
\end{remark}

\begin{remark}
Formulas of Theorem \ref{thm2p1} and
Propositions \ref{prop2p4}, \ref{prop2p5}
are based mainly on asymptotics \eqref{asympsiplus} and, therefore, can be extended to more general discrete Schr\"{o}dinger equations \cite{Eskina2} for which scattering solutions $\psi^+$ are as in \eqref{asympsiplus}.
\end{remark}

\begin{remark}
  A natural open question concerns finding analogues of Theorem \ref{thm2p1} in the case when the condition \eqref{assumeE} is not fulfilled, i.e. when $\left| {{E}}\right|<2d-4.$
  The generalization of asymptotic formula \eqref{asympsiplus}, in this case, is given in \cite{Shaban} and involves several scattered waves with different scattering amplitudes.
  We expect that, for approximately finding these scattering amplitudes, $|\psi^+(x, {k})|^2$ should be measured at several points $x$ and not just two points as in Theorem \ref{thm2p1}.
\end{remark}

\begin{remark}
  To our knowledge, open question also includes establishing the full Atkinson-type expansion for function $\psi^+(x, {k})$ even under conditions \eqref{defv}, \eqref{assumeE}.
  Proceeding from this full expansion, one could develop Theorem \ref{thm2p1} in multi-points' style as, for example, in \cite{Sivkin2}.
\end{remark}

\begin{remark}
  Open questions also include extending Theorem \ref{thm2p1} and formulas \eqref{Dexp}, \eqref{measE} to the case of Schr\"{o}dinger operators on more complicated lattices, for example, as in \cite{Ando}.
\end{remark}

\section{Proofs of Theorem \ref{thm2p1} and formulas \eqref{Dexp} and \eqref{measE}}
\label{proofsec3}

Due to \eqref{asympsiplus},
we have 
\begin{eqnarray}
|\psi^{+}|^2 &=1+e^{-i{{k}}\cdot{{x}}}\dfrac{e^{+i\mu\left(\hat{x}, {{E}}\right) \left| {{x}}\right| }
}{\left| {{x}}\right| ^{\frac{d-1}2}}f^{+}+e^{+i{{k}}\cdot{{x}}}\dfrac{e^{-i\mu\left(\hat{x}, {{E}}\right) \left| {{x}}\right| }}{\left| {{x}}\right|^{\frac{d-1}2}}\overline{f^{+}}\nonumber\\
&\qquad+\frac{1}{\left| {{x}}\right| ^{d-1}}\overline{f^{+}}f^{+}
+O\left(\frac1{\left| {{x}}\right| ^{\frac{d+1}2}}\right) \quad \textrm{ as }|x|\to\infty,
\label{psi2asym}
\end{eqnarray}
where $x\in\mathbb{Z}^d$, $\hat{x}=x/|x|$, $f^{+}=f^{+}(k, \hat{x})$.

Due to smoothness of $f^{+}$, we have that 
\begin{eqnarray}
f^{+}(k, \hat{x})=f^{+}(k, \omega)+O(s^{-1}), \quad f^{+}(k, \hat{y})=f^{+}(k, \omega)+O(s^{-1}),
\label{fpluseq}
\end{eqnarray}
as $s\to+\infty$, where $x$ and $y$ are defined in \eqref{defxy}.
Here we used also formulas \eqref{form2} stated below.

Using \eqref{defaxk}, 
\eqref{psi2asym}, and \eqref{fpluseq}, we obtain
\begin{eqnarray}
e^{-i{{k}}\cdot{{x}}}{e^{+i\mu\left(\hat{x}, {{E}}\right) \left| {{x}}\right| }}f^{+}\left({{k}},\omega\right)
+e^{+i{{k}}\cdot{{x}}}{e^{-i\mu\left(\hat{x}, {{E}}\right) \left| {{x}}\right| }
}\overline{f^{+}}\left({{k}},\omega\right)\nonumber\\\qquad\qquad\qquad\qquad
=a({{x}}, {{k}})+O(s^{-\sigma}),\qquad\label{eq1n}\\
e^{-i{{k}}\cdot{{y}}}{e^{+i\mu\left(\hat{y}, {{E}}\right) \left| {{y}}\right| }
}f^{+}\left({{k}},\omega\right)
+e^{+i{{k}}\cdot{{y}}}{e^{-i\mu\left(\hat{y}, {{E}}\right) \left| {{y}}\right| }
}\overline{f^{+}}\left({{k}},\omega\right)\nonumber\\\qquad\qquad\qquad\qquad
=a({{y}}, {{k}})+O(s^{-\sigma}),\qquad\label{eq2n}
\end{eqnarray}
as $s\to+\infty$, where $x$ and $y$ are defined by \eqref{defxy} and $\sigma$ is as in \eqref{solfplus}.
We consider \eqref{eq1n} and \eqref{eq2n} as a linear system for finding $f^+\left({{k}},\omega\right)$ and $\overline{f^{+}\left({{k}},\omega\right)}$, approximately, in terms of $a({{x}}, {{k}})$ and $a({{y}}, {{k}})$. 
As a result, we get \eqref{solfplus},
where $D$ is the determinant of the $2\times 2$ coefficient matrix.

In order to obtain formula \eqref{Dexp}, we use the definition of $\mu$ in \eqref{defmu}, the definition of $x$ and $y$ in \eqref{defxy}, the definition of $D$ in \eqref{solfplus}, and the formulas:
\begin{eqnarray}
\mu(\hat{x}, {{E}})|x|-\mu(\hat{y}, {{E}})|y|&=&\mu(\hat{x}, {{E}})(|x|-|y|)\nonumber\\
&&-(\mu(\hat{y}, {{E}})-\mu(\hat{x}, {{E}}))|y|,\label{form1a}\\
(\mu(\hat{y}, {{E}})-\mu(\hat{x}, {{E}}))|y|&=&
(\gamma(\hat{y}, E)\cdot\hat{y}-\gamma(\hat{x}, E)\cdot\hat{x})|y|\nonumber\\
&=&(\gamma(\hat{y}, E)-\gamma(\hat{x}, E))\cdot\hat{y}|y|\nonumber\\
&&+\gamma(\hat{x}, E)\cdot(\hat{y}-\hat{x})|y|;\qquad
\label{form2a}
\end{eqnarray}

\begin{eqnarray}
|y|&=&|x+\zeta|=\sqrt{|x|^2+2x\cdot\zeta+|\zeta|^2}\nonumber\\ \qquad
&=&|x|+\hat{x}\cdot\zeta+O(|x|^{-1}) \textrm{ as }|x|\to\infty,
\label{form1}\\
\hat{x}&=&\omega+O(s^{-1}), \quad\hat{y}=\omega+O(s^{-1}) \textrm{ as }s\to+\infty.
\label{form2}
\end{eqnarray}
Using \eqref{form1}, \eqref{form2}, and smoothness of $\mu$, we get
\begin{eqnarray}
\mu(\hat{x}, E)(|x|-|y|)
=-\mu(\omega, E)\omega\cdot\zeta+O(s^{-1})\textrm{ as }s\to+\infty.
\label{formeqna}
\end{eqnarray}
In order to estimate $(\mu(\hat{y}, {{E}})-\mu(\hat{x}, {{E}}))|y|$, we use formulas
\eqref{defmu},
\eqref{form2a}-\eqref{form2}
and also the formulas 
\begin{eqnarray}
\gamma(\hat{x}, E)-\gamma(\omega, E)&=&O(s^{-1}), \quad\gamma(\hat{y}, E)-\gamma(\omega, E)=O(s^{-1}),\qquad\label{addform0}\\
(\gamma(\hat{y}, E)-\gamma(\hat{x}, E))\cdot\omega
&=&O(s^{-2}),\label{addform1}\\
\hat{y}-\hat{x}&=&s^{-1}\Pi_{\omega}\zeta +O(s^{-2}),
\label{addform2}
\end{eqnarray}
as $s\to+\infty$, where 
\begin{eqnarray}
\Pi_{\omega}\zeta:=\zeta-(\omega\cdot\zeta)\omega.
\label{Piform}
\end{eqnarray}

As a result, using also smoothness of $\gamma,$
we get
\begin{eqnarray}
(\mu(\hat{y}, {{E}})-\mu(\hat{x}, {{E}}))|y|&=\gamma(\omega, E)\cdot\zeta-\gamma(\omega, E)\cdot\omega(\omega\cdot\zeta)+O(s^{-1}).
\label{finaleqn}
\end{eqnarray}

Note that  to get formula \eqref{addform1} we use formulas \eqref{addform0} and that $\gamma(\omega, E)$, $\gamma(\hat{x}, E)$, $\gamma(\hat{y}, E)\in\Gamma(E)$,
where the normal to $\Gamma(E)$ at the point $\gamma(\omega, E)$ coincides with $\omega$, and that $\Gamma(E)$ is smooth.
Therefore, 
$(\gamma-\gamma(\omega, E))\cdot\omega=O(|\gamma-\gamma(\omega, E)|^2)$ for $\gamma=\gamma(\hat{x}, E)$ or $\gamma=\gamma(\hat{y}, E)$.
And we use also that $\gamma(\hat{y}, E)-\gamma(\hat{x}, E)=(\gamma(\hat{y}, E)-\gamma(\omega, E))
+(\gamma(\omega, E)-\gamma(\hat{x}, E)).$

From \eqref{form1a}, \eqref{formeqna}, and \eqref{finaleqn}, we get
\begin{eqnarray}
\mu(\hat{x}, {{E}})|x|-\mu(\hat{y}, {{E}})|y|
=-\gamma(\omega, E)\cdot\zeta+O(s^{-1}).
\label{finaleqn2}
\end{eqnarray}
Substituting \eqref{finaleqn2}
in the definition of  $D$ in \eqref{solfplus}, we obtain the stated result \eqref{Dexp}.

Under assumption \eqref{assumeE}, formula \eqref{measE} follows from the statements:\\
(i) The function $\alpha(\omega):=({{k}}-\gamma(\omega, {{E}}))\cdot\zeta$ is real-analytic in $\omega\in S^{d-1}$ for fixed ${k}$ and $\zeta$.\\
(ii) $\alpha(\omega)$ is not identically constant.\\
(iii) $\textrm{Meas }\mathcal{E}=0$ in $S^{d-1}$ if $\mathcal{E}$ is the set of zeroes of a non-zero real-analytic function $u$ on $S^{d-1}$.

Statement (i) follows from 
the real-analyticity of ${\gamma}$ in formula \eqref{defkout}.
The latter analyticity of ${\gamma}$ follows from analyticity, strict convexity, and non-zero principal curvatures of $\Gamma(E).$
Alternatively, one can use explicit formulas, relating $\omega$ and $\gamma$, mentioned in the Proof of Lemma 3 in \cite{Shaban} (where $\gamma$ is denoted as $k$ and $E$ is denoted as $\lambda$).

Statement (ii) follows from the property that $\gamma(\omega, E)\cdot\zeta$ is not identically constant. 
The reason is that $\Gamma(E)$ does not belong to a hyper-plane in $\mathbb{R}^d$.

Statements of type statement (iii) are well-known in analysis; see, for example, \cite{Mityagin} and references therein.

Actually, formula \eqref{defEkg} follows from statement (iii) used for $u=\alpha-n\pi$ for several integer $n$. 

\section{Proof of Proposition \ref{prop2p4n}}
\label{proofsec4n}
Using formulas \eqref{psi2asym}, \eqref{fpluseq},
in a way similar to formulas \eqref{eq1n}, \eqref{eq2n},
we have also that
\begin{eqnarray}
e^{-i{{k}}\cdot{{x}}}{e^{+i\mu\left(\hat{x}, {{E}}\right) \left| {{x}}\right| }}f^{+}
+e^{+i{{k}}\cdot{{x}}}{e^{-i\mu\left(\hat{x}, {{E}}\right) \left| {{x}}\right| }
}\overline{f^{+}}
=a({{x}}, {{k}})-\frac{1}{s^{1/2}}\overline{f^{+}}f^{+}+O(s^{-1}),\qquad\label{eq1nd2}\\
e^{-i{{k}}\cdot{{y}}}{e^{+i\mu\left(\hat{y}, {{E}}\right) \left| {{y}}\right| }
}f^{+}
+e^{+i{{k}}\cdot{{y}}}{e^{-i\mu\left(\hat{y}, {{E}}\right) \left| {{y}}\right| }
}\overline{f^{+}}
=a({{y}}, {{k}})-\frac{1}{s^{1/2}}\overline{f^{+}}f^{+}+O(s^{-1}),\qquad\label{eq2nd2}
\end{eqnarray}
as $s\to+\infty$, where $x$ and $y$ are defined by \eqref{defxy}, and $f^{+}=f^{+}(k, \omega)$.

In addition, using formulas \eqref{solfplus} and \eqref{deff0plus},
we have that
\begin{eqnarray}
\overline{f^{+}}f^{+}
&=&(\overline{f^{+}_0}+\frac{1}{D}O(s^{-1/2}))f^{+}
=\overline{f^{+}_0}f^{+}+\frac{1}{D}O(s^{-1/2})\nonumber\\
&=&\overline{f^{+}_0}(f^{+}_0+\frac{1}{D}O(s^{-1/2}))+\frac{1}{D}O(s^{-1/2})\nonumber\\
&=&\overline{f^{+}_0}f^{+}_0+\frac{1}{D^2}O(s^{-1/2}), \textrm{ as }s\to+\infty.
\label{ffplusnew}
\end{eqnarray}
Substituting  \eqref{ffplusnew}  into \eqref{eq1nd2}, \eqref{eq2nd2}, we obtain
\begin{eqnarray}
e^{-i{{k}}\cdot{{x}}}{e^{+i\mu\left(\hat{x}, {{E}}\right) \left| {{x}}\right| }}f^{+}
+e^{+i{{k}}\cdot{{x}}}{e^{-i\mu\left(\hat{x}, {{E}}\right) \left| {{x}}\right| }
}\overline{f^{+}}\nonumber\\\qquad\qquad\qquad\qquad
=a({{x}}, {{k}})-\frac{1}{s^{1/2}}\overline{f^{+}_0}f^{+}_0+\frac{1}{D^2}O(s^{-1}),\qquad\label{eq1nd2s}\\
e^{-i{{k}}\cdot{{y}}}{e^{+i\mu\left(\hat{y}, {{E}}\right) \left| {{y}}\right| }
}f^{+}
+e^{+i{{k}}\cdot{{y}}}{e^{-i\mu\left(\hat{y}, {{E}}\right) \left| {{y}}\right| }
}\overline{f^{+}}\nonumber\\\qquad\qquad\qquad\qquad
=a({{y}}, {{k}})-\frac{1}{s^{1/2}}\overline{f^{+}_0}f^{+}_0+\frac{1}{D^2}O(s^{-1}),\qquad\label{eq2nd2s}
\end{eqnarray}
as $s\to+\infty,$ where $f^{+}_0=f^{+}_0(k, \omega)$.

Solving the system \eqref{eq1nd2s}, \eqref{eq2nd2s} with respect to $f^{+}$, we get
\begin{eqnarray}f^+(k, \omega)
&=&\frac{1}{D}\bigg(e^{i({{k}}\cdot{{y}}-\mu\left(\hat{y}, {{E}}\right) \left| y\right| )} a(x, {k}) - e^{i({{k}}\cdot{{x}}-\mu\left(\hat{x}, {{E}}\right) \left| x\right| )}a(y, {k})\nonumber\\
&&-(e^{i({{k}}\cdot{{y}}-\mu\left(\hat{y}, {{E}}\right) \left| y\right| )}-e^{i({{k}}\cdot{{x}}-\mu\left(\hat{x}, {{E}}\right) \left| x\right| )})\frac{1}{s^{1/2}}\overline{f^{+}_0}f^{+}_0+\frac{1}{D^2}O(s^{-1})\bigg)\nonumber\\
&=&f^+_0(k, \omega)-\frac{1}{D}(e^{i({{k}}\cdot{{y}}-\mu\left(\hat{y}, {{E}}\right) \left| y\right| )}-e^{i({{k}}\cdot{{x}}-\mu\left(\hat{x}, {{E}}\right) \left| x\right| )})\frac{1}{s^{1/2}}\overline{f^{+}_0}f^{+}_0\nonumber\\
&&+\frac{1}{D^3}O(s^{-1})\textrm{ as }s\to+\infty.
\end{eqnarray}

This completes the proof.

\section{Proofs of Propositions \ref{prop2p4} and \ref{prop2p5}}
\label{proofsec4}
Due to \eqref{asympsiplus} and \eqref{eqnk1d}, with $O\left(\left| {{x}}\right| ^{-\frac{d+1}2}\right)\equiv0$ for $d=1$,
we have
\begin{eqnarray}
s_{21}e^{-2ikx}+\overline{s_{21}}e^{2ikx}+|s_{21}|^2=a(x, k).
\label{eq1proof2p4}
\end{eqnarray}
To prove Proposition \ref{prop2p4}, we use \eqref{eq1proof2p4}
and also \eqref{eq1proof2p4} with $x$ replaced by $y$, i.e., 
\begin{eqnarray}
s_{21}e^{-2iky}+\overline{s_{21}}e^{2iky}+|s_{21}|^2=a(y, k).
\label{eq2proof2p4}
\end{eqnarray}
As a result we consider \eqref{eq1proof2p4} and \eqref{eq2proof2p4} as a linear system for finding $s_{21}$ and $\overline{s_{21}}$ from $a(x, k)-|s_{21}|^2$
and $a(y, k)-|s_{21}|^2$.
Solving this system, we get formula \eqref{eqprop2p4}.

To prove Proposition \ref{prop2p5}, we consider formula \eqref{eq1proof2p4} with $x=x_1, x_2, x_3$.
Subtracting equality \eqref{eq1proof2p4} for $x=x_1$ from equality \eqref{eq1proof2p4} for $x=x_2$ and from equality \eqref{eq1proof2p4} for $x=x_3$, we obtain the system
\begin{eqnarray}
s_{21}(e^{-2ikx_2}-e^{-2ikx_1})+\overline{s_{21}}(e^{2ikx_2}-e^{2ikx_1})=a(x_2, {k})-a(x_1, {k}),\label{eq2aproof2p5}\\
s_{21}(e^{-2ikx_3}-e^{-2ikx_1})+\overline{s_{21}}(e^{2ikx_3}-e^{2ikx_1})=a(x_3, {k})-a(x_1, {k}),\label{eq2bproof2p5}
\end{eqnarray}
for finding $s_{21}$ and $\overline{s_{21}}$.
One can see that 
\begin{eqnarray}
D=2i\bigg(\sin(2k(x_3 - x_2)) +\sin(2k(x_2 - x_1)) + \sin(2k(x_1 - x_3))\bigg),
\end{eqnarray}
where $D$ is the determinant of the system \eqref{eq2aproof2p5}, \eqref{eq2bproof2p5}.
This determinant can be also re-written in the form of $D$ in \eqref{eqprop2p5}; see (3.7) in the arXiv preprint of \cite{Novikov1d}.
Solving \eqref{eq2aproof2p5}, \eqref{eq2bproof2p5} we get formula \eqref{eqprop2p5}.
Due to our assumption that $x_i\ne x_j\textrm{ mod }(\pi/{k})$, we have that $D\ne0$ in \eqref{eqprop2p5}.

\section*{Acknowledgments}
This work was started during the stay (in May 2023) of both authors at the Isaac Newton Institute (INI) for Mathematical Sciences, Cambridge.
The authors would like to thank 
INI
for support and hospitality during the programmes -- `{\em Mathematical theory and applications of multiple wave scattering}'
(MWS)
and
`{\em Rich and Nonlinear Tomography - a multidisciplinary approach}' (RNT) -- where work on this paper was undertaken. This work was supported by EPSRC grant no EP/R014604/1. 
A part of the work of BLS, for the same visit to INI during January-June 2023 and participation in MWS, was partially supported by a grant from the Simons Foundation.

\end{document}